\documentclass[useAMS,usenatbib]{mn2e}

\usepackage{graphicx}
\newcommand{\eq}[1]{\begin{equation}  #1 \end{equation}}

\newcommand{\eqa}[1]{\begin{eqnarray}   #1 \end{eqnarray}}

\newcommand{\br}[1]{\left( #1 \right)}
\newcommand{\bc}[1]{\left\{ #1 \right\}}
\newcommand{\bb}[1]{\left[ #1 \right]}
\newcommand{\ba}[1]{\left\langle #1 \right\rangle}

\newcommand{\nn}{\nonumber}

\newcommand{\dd}{{\rm d}}

\newcommand{\vek}[1]{\mbox{\boldmath $#1$}}

\title{Forecasts of non-Gaussian parameter spaces using Box-Cox transformations}
\author[B. Joachimi \& A.N. Taylor]
  {B.~Joachimi$^1$\thanks{E-mail: bj@roe.ac.uk}
  and A.N.~Taylor$^1$\\
  $^1$Institute for Astronomy, University of Edinburgh, Royal Observatory, Blackford Hill, Edinburgh, EH9 3HJ, U.K.}
\date{Accepted . Received ; in original form }

\pagerange{\pageref{firstpage}--\pageref{lastpage}} \pubyear{2011}

\begin{document}
\label{firstpage}

\maketitle

\begin{abstract}
Forecasts of statistical constraints on model parameters using the Fisher matrix abound in many fields of astrophysics. The Fisher matrix formalism involves the assumption of Gaussianity in parameter space and hence fails to predict complex features of posterior probability distributions. Combining the standard Fisher matrix with Box-Cox transformations, we propose a novel method that accurately predicts arbitrary posterior shapes. The Box-Cox transformations are applied to parameter space to render it approximately multivariate Gaussian, performing the Fisher matrix calculation on the transformed parameters. We demonstrate that, after the Box-Cox parameters have been determined from an initial likelihood evaluation, the method correctly predicts changes in the posterior when varying various parameters of the experimental setup and the data analysis, with marginally higher computational cost than a standard Fisher matrix calculation. We apply the Box-Cox-Fisher formalism to forecast cosmological parameter constraints by future weak gravitational lensing surveys. The characteristic non-linear degeneracy between matter density parameter and normalisation of matter density fluctuations is reproduced for several cases, and the capabilities of breaking this degeneracy by weak lensing three-point statistics is investigated. Possible applications of Box-Cox transformations of posterior distributions are discussed, including the prospects for performing statistical data analysis steps in the transformed Gaussianised parameter space.
\end{abstract}


\section{Introduction}

In recent years many fields of astrophysics have seen a transition towards increasingly large experiments and surveys. The level of complexity and the costs are rising alongside, requiring careful planning and assessment of the expected performance of the envisaged project at all stages. In forecasts of the statistical constraints on model parameters by future experiments the Fisher matrix \citep{fisher35,tegmark97} has proven to be indispensable \citep[e.g.][]{albrecht06}.

Its ubiquity can largely be attributed to the low computational cost of a Fisher matrix calculation compared to a full mock likelihood analysis, in particular if the data set to be analysed and the number of parameters to be inferred are large. This simplicity comes at the price of a twofold assumption of Gaussianity in the derivation of the Fisher matrix expressions \citep{tegmark97}. First, although not a requirement of the Fisher matrix formalism, the data are usually assumed to be distributed according to a multivariate Gaussian. This assumption is shared with the majority of full likelihood analyses to date, invoking the central limit theorem or simply because the precise distribution is unknown or intractable, but precision measurements may eventually require more complicated forms \citep[e.g.][]{bond00,hartlap09}. 

Second, since the Fisher matrix is defined as the expectation value of the Hessian of the log-likelihood in parameter space, it describes the shape of the posterior distribution up to second order locally in the vicinity of the maximum likelihood point. Hence it can only provide an accurate global representation of a posterior whose logarithm has constant curvature, i.e. is Gaussian.

Therefore the confidence levels on model parameters derived from a Fisher matrix are inevitably elliptical. They can describe the posterior distribution close to the point of maximum likelihood and indicate linear degeneracies among parameters via the ellipticity of confidence regions. Fisher matrix analyses fail to identify the shape of the posterior away from its maximum, as well as to detect non-linear dependencies of parameters. However, non-linear model parameter degeneracies are common, and the attempt to minimise or break them can drive the design of experiments. Hence it is desirable to go beyond the assumption of a Gaussian posterior in forecasts for the advanced stages of upcoming precision measurements.

In this work we propose to combine Fisher matrix forecasts with Box-Cox transformations of parameter space to obtain accurate expectations of posterior distributions. \citet{box64} introduced a parametrised set of power transformations with which a wide range of data can be transformed to follow a Gaussian distribution to good approximation. We will apply these transformations to model parameters, instead of data, in order to modify a given posterior into a multivariate Gaussian distribution for which a Fisher matrix analysis is exact. After an inverse Box-Cox transformation the Fisher matrix results will then accurately describe the original posterior. To determine the free parameters of the Box-Cox transformation, the original posterior needs to be sampled, and hence an initial mock likelihood analysis to be run.

We will demonstrate this method with an example from cosmology. Several ambitious surveys are currently planned or designed\footnote{These include e.g. the Large Synoptic Survey Telescope (\texttt{http://www.lsst.org}), the NASA satellite WFIRST (\texttt{http://wfirst.gsfc.nasa.gov}), and the ESA satellite Euclid (\texttt{http://sci.esa.int/euclid}).} that are going to measure the parameters of the cosmological standard model, particularly those of dark matter and dark energy, with high precision. These experiments will investigate several cosmological probes of the large-scale structure of the Universe, the potentially most powerful one being weak gravitational lensing of distant galaxies \citep{albrecht06,peacock06}. Weak lensing features a characteristic non-linear degeneracy between the two best-constrained parameters $\Omega_{\rm m}$ (mean matter density) and $\sigma_8$ (normalisation of matter density fluctuations) as they both govern the overall amplitude of the signal \citep[see e.g.][]{hoekstra06,schrabback09}. Hence a mock weak lensing survey provides an excellent test case, but we emphasise that the method outlined is applicable to any prediction for model parameter constraints.

The paper is organised as follows. Section \ref{sec:theory} details the principles of Box-Cox transformations, our strategies to determine optimal Box-Cox parameters, and the combined Box-Cox-Fisher formalism. In Section \ref{sec:performance} we investigate the performance of the proposed method for a mock weak lensing experiment, comparing different variants in the implementation and quantifying the universality of the Box-Cox-Fisher formalism. We apply this formalism to a test of the degeneracy-breaking capabilities of weak lensing higher-order statistics in Section \ref{sec:application}, before we summarise and conclude on our findings in Section \ref{sec:conclusions}.

\section{Box-Cox transformations of parameter space}
\label{sec:theory}

Power transformations such as the inverse and square-root transformation, or logarithmic transformations are popular choices to render the distribution of data more Gaussian. The Box-Cox transformation unites these cases with a single free parameter per dimension and are hence widely used in various areas of science. Astrophysical applications are rare; one example is the work by \citet{dineen05} who tested cosmic microwave background data for Gaussianity.

For a $N_p$-dimensional variable $\vek{p}$ the Box-Cox transformation in each dimension $\mu=1,\,..\,,N_p$ reads \citep{box64}
\eq{
\label{eq:bctrafo}
\bar{p}_\mu(\lambda_\mu,a_\mu) = \left\{ \begin{array}{ll} \bb{\br{p_\mu + a_\mu}^{\lambda_\mu}-1}/\lambda_\mu & \lambda_\mu \neq 0\\ \ln (p_\mu + a_\mu) & \lambda_\mu = 0 \end{array} \right. \;,
}
where the normalisation has been chosen such that the transformation is continuous in the parameter $\lambda_\mu$ at $\lambda_\mu=0$. We allow for a shift $a_\mu$ as a second free parameter in each dimension. Note that we denote transformed quantities by a bar and drop the dependence on the Box-Cox parameters $\br{\vek{\lambda},\vek{a}}$ unless it needs to be made explicit. 

Usually, equation (\ref{eq:bctrafo}) is applied to the elements of a data-vector, but we will henceforth understand $p_\mu$ as the parameters of an $N_p$-dimensional parameter space. Then the transform of a given posterior distribution ${\cal P}(\vek{p})$ is given by
\eq{
\label{eq:bcdistribution}
\bar{\cal P}(\bar{\vek{p}}) = {\cal P}(\vek{p})\; J(\vek{p},\bar{\vek{p}})\;,
}
with the Jacobian 
\eq{
\label{eq:jacobian}
J(\vek{p},\bar{\vek{p}}) = \prod_{\mu=1}^{N_p} \left| \frac{\dd p_\mu}{\dd \bar{p}_\mu} \right| = \prod_{\mu=1}^{N_p} \br{p_\mu + a_\mu}^{1-\lambda_\mu}\;.
}
The second equality follows directly from equation (\ref{eq:bctrafo}). The first goal is to determine the set of $2 N_p$ parameters, $\br{\vek{\lambda},\vek{a}}$, such that the transformed posterior, $\bar{\cal P}(\bar{\vek{p}})$, is a multivariate Gaussian to good approximation.

\subsection{Optimal Box-Cox parameters}
\label{sec:optimalbc}

Suppose a random sample $\vek{\hat{p}}$ with $n$ elements, i.e. $\bc{\hat{p}_{\mu,1},\,..\,,\hat{p}_{\mu,n}}$ for every $\mu=1,\,..\,,N_p$, is drawn from the posterior ${\cal P}(\vek{p})$, for instance via Monte-Carlo sampling techniques. If the Box-Cox transformed posterior is indeed Gaussian, the distribution is given by
\eqa{
\label{eq:transformedposterior}
{\cal P}(\vek{p}) &=& \bar{\cal P}(\bar{\vek{p}})\; J(\bar{\vek{p}},\vek{p})\\ \nn
&=& \prod_{\mu=1}^{N_p} \br{p_\mu + a_\mu}^{\lambda_\mu-1}\; \frac{1}{\sqrt{(2\pi)^{N_p}\det {\rm Cov}(\bar{\vek{p}})}}\\ \nn
&& \times\; \exp \bc{- \frac{1}{2}\; (\bar{\vek{p}}-\bar{\vek{p}}_{\rm max})^\tau\; {\rm Cov}^{-1}(\bar{\vek{p}})\; (\bar{\vek{p}}-\bar{\vek{p}}_{\rm max})}
}
and has only the Box-Cox parameters, and the mean $\bar{\vek{p}}_{\rm max}$ and covariance
\eq{
\label{eq:defcov}
{\rm Cov}(\bar{\vek{p}}) \equiv \ba{(\bar{\vek{p}}-\bar{\vek{p}}_{\rm max})\;(\bar{\vek{p}}-\bar{\vek{p}}_{\rm max})^\tau}\;
} 
of the Gaussian as free parameters. Since $\bar{\cal P}(\bar{\vek{p}})$ is assumed Gaussian, one can employ the standard maximum likelihood estimators for the covariance and mean. The latter simply implies $\bar{\vek{p}}=\bar{\vek{p}}_{\rm max}$, so that the exponential in (\ref{eq:transformedposterior}) is unity. Consequently one obtains the following concentrated log-likelihood for the Box-Cox parameters \citep[for details see][]{box64,velilla93},
\eqa{
\label{eq:bclikelihood}
{\cal L}_{\rm max}(\vek{\lambda},\vek{a}) &=& - \frac{n}{2}\; \ln \det {\rm Cov}\bb{\bar{\vek{\hat{p}}}(\vek{\lambda},\vek{a})}_{\rm ML}\\ \nn
&& + \sum_{\mu=1}^{N_p} \bc{(\lambda_\mu -1) \sum_{i=1}^n \ln (\hat{p}_{\mu,i}+a_\mu)}\;,
}
up to an irrelevant constant. We have added the subscript ML to emphasise that the maximum likelihood estimate for the covariance based on $\vek{\hat{p}}$ is to be used. Maximising this likelihood for a given sample $\vek{\hat{p}}$ should then return Box-Cox parameters $\br{\vek{\lambda},\vek{a}}$ that render $\bar{\cal P}(\bar{\vek{p}})$ as close to Gaussian as possible.

If $N_p$ is small or the likelihood evaluation computationally inexpensive, it may be more convenient and faster to obtain the distribution ${\cal P}(\vek{p})$ directly on a grid instead of using a random sample \citep[see also][]{frommert10}. In this case the transformed posterior can be computed readily via equation (\ref{eq:bcdistribution}) for any combination of Box-Cox parameters. The optimal parameter combination is then found by comparing $\bar{\cal P}(\bar{\vek{p}})$ to a Gaussian distribution with the same mean and covariance, e.g. by minimising the Kullback-Leibler divergence
\eqa{
\label{eq:kullbackleibler}
D_{\rm KL} &=& \int \dd^{N_p} p\; {\cal P}_{\rm ref}(\vek{p}) \ln \frac{{\cal P}_{\rm ref}(\vek{p})}{{\cal P}(\vek{p})}\\ \nn
&\approx& \sum_i {\cal P}_{\rm ref}(\vek{p}_i) \ln \frac{{\cal P}_{\rm ref}(\vek{p}_i)}{{\cal P}(\vek{p}_i)} \prod_{\mu=1}^{N_p} \Delta p_\mu\;.
}
In the second equality we have replaced the integration with a sum over all points of the grid on which the distributions are evaluated, assuming a spacing of the points by $\Delta p_\mu$ in dimension $\mu$. However, we will use $D_{\rm KL}$ to assess the accuracy of the results of our method in Section \ref{sec:performance}, so that we use a different statistic to determine the Box-Cox parameters.

Two one-dimensional distributions can be compared via their quantiles in a QQ-plot. If both distributions are Gaussian, the quantile pairs lie on a straight and hence Pearson's correlation coefficient of the quantiles,
\eq{
\label{eq:rqq}
r_{\rm QQ} = \frac{\ba{\br{Q^{\rm trans}-\langle Q^{\rm trans} \rangle} \bigl(Q^{\rm gauss}-\langle Q^{\rm gauss} \rangle \bigr)}}{\sqrt{\ba{\br{Q^{\rm trans}-\ba{Q^{\rm trans}}}^2}\; \ba{\br{Q^{\rm gauss}-\ba{Q^{\rm gauss}}}^2}}}\;,
}
should attain unity. Here, $Q^{\rm trans}$ denotes the quantiles of the Box-Cox transformed distribution and $Q^{\rm gauss}$ the quantiles of a zero-mean unit-variance Gaussian distribution, the latter readily computed from the cumulative distribution function. In practice we use 30-quantiles to calculate equation (\ref{eq:rqq}). An advantage of $r_{\rm QQ}$ over $D_{\rm KL}$ is that it is independent of the mean and variance of the transformed distribution which therefore do not have to be re-computed for every change in Box-Cox parameters.

Since $r_{\rm QQ}$ can only be applied to one-dimensional distributions, we determine the Box-Cox parameters in every dimension of parameter space from the marginalised posterior in that dimension. When following the approach of using a random sample $\vek{\hat{p}}$ together with equation (\ref{eq:bclikelihood}) to optimise the Box-Cox parameters, we will compare the performance of determining $\br{\vek{\lambda},\vek{a}}$ from the full $N_p$-dimensional posterior and the $N_p$ marginal posteriors.

\subsection{Box-Cox-Fisher formalism}
\label{sec:bcfisher}

Once the optimal Box-Cox parameters are found by either of the methods described in the foregoing section, one can proceed to unite the Box-Cox transformations with the Fisher matrix technique. If the same set of experimental parameters is used for the Box-Cox-Fisher prediction as for the fiducial mock likelihood analysis that the optimal Box-Cox parameters were determined from, one should obtain identical results. Changing the experimental setup in the Box-Cox-Fisher forecasts should then yield similarly accurate results, as long as these parameters do not depart too strongly from those of the mock likelihood analysis such that the shape of the posterior would be modified significantly. The universality with respect to changes in various parameters of the exemplary weak lensing survey will be tested in Section \ref{sec:performance}. 

The task is hence to compute the posterior distribution, ${\cal P}(\vek{p})$, of model parameters $\vek{p}$ for a given set of Box-Cox parameters $\br{\vek{\lambda},\vek{a}}$ and a standard Fisher matrix $F^{\rm orig}$, computed for at a fiducial point $\vek{p}_{\rm fid}$ in parameter space. In analogy to equation (\ref{eq:transformedposterior}) the posterior is given by
\eqa{
\label{eq:transformedposteriorfisher}
\nn
{\cal P}(\vek{p}) &=& \!\!\! \sqrt{\frac{\det \bar{F}}{(2\pi)^{N_p}}}\; \exp \bc{- \frac{1}{2}\; (\bar{\vek{p}}-\bar{\vek{p}}_{\rm max})^\tau\; \bar{F}\; (\bar{\vek{p}}-\bar{\vek{p}}_{\rm max})}\\ 
&& \times\; \prod_{\mu=1}^{N_p} \br{p_\mu + a_\mu}^{\lambda_\mu-1}\;,
}
where we used the transformed Fisher matrix $\bar{F}$ as an estimator for the inverse covariance of the Gaussian of the Box-Cox transformed posterior. The peak position $\bar{\vek{p}}_{\rm max}$ of this Gaussian and $\bar{F}$ are the only unknown quantities in equation (\ref{eq:transformedposteriorfisher}) that have yet to be determined. 

In the following we will assume, as in the standard derivation of the Fisher matrix, that the prior is uniform in the region of parameter space where the likelihood deviates significantly from zero. Thus the log-likelihood is given by ${\cal L} = - \ln {\cal P}$, and likewise for the transformed posterior. Then equation (\ref{eq:transformedposteriorfisher}) is equivalent to
\eq{
\label{eq:loglike}
\bar{\cal L} = {\cal L} - \sum_{\mu=1}^{N_p} (1-\lambda_\mu)\; \ln \br{p_\mu + a_\mu}\;.
}
If we designate $\vek{p}_{\rm max}$ as the result of an inverse Box-Cox transformation of $\bar{\vek{p}}_{\rm max}$ and employ the definition of the Fisher matrix, we arrive at the following expression for the transformed Fisher matrix,
\eqa{
\label{eq:bcfisher1}
&& \bar{F}_{\mu\nu} \equiv \ba{ \left. \frac{\partial^2 \bar{\cal L}}{\partial \bar{p}_\mu \partial \bar{p}_\nu} \right|_{\bar{\vek{p}}_{\rm max}}}\\ \nn
&=& \!\!\! \ba{ \left. \frac{\partial^2 \cal L}{\partial p_\mu \partial p_\nu} \right|_{\vek{p}_{\rm max}}} \br{p_{\mu,{\rm max}}+a_\mu}^{1-\lambda_\mu} \br{p_{\nu,{\rm max}}+a_\nu}^{1-\lambda_\nu}\\ \nn
&& +\; \delta_{\mu\nu} \Biggl\{ \lambda_\mu (\lambda_\mu-1) \br{p_{\mu,{\rm max}}+a_\mu}^{-2\lambda_\mu}\\ \nn
&& + (1-\lambda_\mu) \br{p_{\mu,{\rm max}}+a_\mu}^{1-2\lambda_\mu} \ba{ \left. \frac{\partial \cal L}{\partial p_\mu} \right|_{\vek{p}_{\rm max}}} \Biggr\}\;.
}
Here, angular brackets denote expectation values, and $\delta_{\mu\nu}$ is the Kronecker symbol.

At this point we make the simplifying assumption that $\vek{p}_{\rm max} \approx \vek{p}_{\rm fid}$, i.e. that the Box-Cox transformation maps the peak of the original posterior onto the peak of the transformed posterior. As will be demonstrated below, this approximation holds to high accuracy. Alternatively, one could instead Taylor-expand the expectation values of the first and second derivatives of ${\cal L}$ in equation (\ref{eq:bcfisher1}), but this step would necessitate the computation of third-order derivatives of ${\cal L}$ already at the first order of the expansion.

Replacing $\vek{p}_{\rm max}$ by $\vek{p}_{\rm fid}$ in equation (\ref{eq:bcfisher1}), the expectation of the first derivative of the log-likelihood vanishes because it has a maximum at $\vek{p}_{\rm fid}$. Invoking the definition of the standard Fisher matrix for the original distribution, one obtains
\eqa{
\label{eq:bcfisher2}
\bar{F}_{\mu\nu} &\approx& F_{\mu\nu}^{\rm orig}\; \br{p_{\mu,{\rm fid}}+a_\mu}^{1-\lambda_\mu} \br{p_{\nu,{\rm fid}}+a_\nu}^{1-\lambda_\nu}\\ \nn
&& +\; \delta_{\mu\nu}\; \lambda_\mu\; (\lambda_\mu-1) \br{p_{\mu,{\rm fid}}+a_\mu}^{-2\lambda_\mu}\;.
}

We pursue two approaches to determine $\bar{\vek{p}}_{\rm max}$, or equivalently, $\vek{p}_{\rm max}$. Requiring that the transformed posterior peaks at $\bar{\vek{p}}_{\rm max}$ yields the condition
\eqa{
\label{eq:pmaxforward}
\ba{ \left. \frac{\partial \bar{\cal L}}{\partial \bar{p}_\mu} \right|_{\bar{\vek{p}}_{\rm max}}} &=& \ba{ \left. \frac{\partial \cal L}{\partial p_\mu} \right|_{\vek{p}_{\rm max}}} \br{p_{\mu,{\rm max}}+a_\mu}^{1-\lambda_\mu}\\ \nn
&& \;+ (\lambda_\mu-1) \br{p_{\mu,{\rm max}}+a_\mu}^{-\lambda_\mu} = 0\;,
}
which can be numerically solved after Taylor-expanding the expectation value around $\vek{p}_{\rm fid}$,
\eq{
\label{eq:taylordl}
\ba{ \left. \frac{\partial \cal L}{\partial p_\mu} \right|_{\vek{p}_{\rm max}}} \approx \sum_{\nu=1}^{N_p} F_{\mu\nu}^{\rm orig} \br{p_{\nu,{\rm max}}-p_{\nu,{\rm fid}}}\;.
}
Alternatively, one can determine $\vek{p}_{\rm max}$ such that the original distribution ${\cal P}(\vek{p})$ peaks at $\vek{p}_{\rm fid}$, which, using equation (\ref{eq:loglike}), leads to the condition
\eqa{
\label{eq:pmaxbackward}
\ba{ \left. \frac{\partial {\cal L}}{\partial p_\mu} \right|_{\vek{p}_{\rm fid}}} &=& \br{p_{\mu,{\rm fid}}+a_\mu}^{\lambda_\mu-1}\\ \nn
&& \hspace*{-2.7cm} \times\; \sum_{\nu=1}^{N_p} \bar{F}_{\mu\nu} \br{\bar{p}_{\nu,{\rm fid}} - \bar{p}_{\nu,{\rm max}}} - (\lambda_\mu-1) \br{p_{\mu,{\rm fid}}+a_\mu}^{-1} = 0\;.
}
After inserting the approximation given by equation (\ref{eq:bcfisher2}), one obtains an expression that can analytically be solved for $\vek{p}_{\rm max}$. Since both procedures involve approximations, we will compare their performance below in Section \ref{sec:comparison}.

Gaussian priors can be added to the diagonal of $F_{\mu\nu}^{\rm orig}$ in the same way as for the standard Fisher analysis, but if the priors modify the posterior substantially, they also have to be included in the mock likelihood analysis used to find optimal Box-Cox parameters. Note that, when grid or Monte-Carlo sampling this likelihood, one usually defines a maximum range in which the model parameters are allowed to vary. This corresponds to an implicit top-hat prior which cannot be represented in the Fisher matrix formalism. Hence, one has to make sure that the posterior used to determine Box-Cox parameters lies well within the parameter space considered.

\section{Performance}
\label{sec:performance}

To assess the performance of Fisher matrix forecasts combined with Box-Cox transformations, we consider a mock weak lensing survey as outlined in the following. While the modelling is at a level of realism similar to current predictions for planned observational projects, we do not attempt to mimic any particular survey, but rather choose the survey characteristics such that we obtain a posterior distribution of cosmological parameters which serves as a particularly useful benchmark. 

Hence, our mock survey will produce a pronounced non-linear degeneracy between the parameters $\Omega_{\rm m}$, the matter density, and $\sigma_8$, the normalisation of matter density fluctuations as an ideal test case. Note that actual future weak lensing surveys will generate much stronger parameter constraints and a reduced $\Omega_{\rm m}-\sigma_8$ degeneracy, so that the Box-Cox-Fisher formalism should perform well in these cases once it does so for the scenario studied in this work.

We will then investigate in detail the implementation outlined in Section \ref{sec:theory}, before answering the question how accurate the Box-Cox-Fisher formalism is when varying the fiducial cosmology, survey parameters, the weak lensing statistic entering the likelihood, and the dimension of the posterior distribution. To be of practical use, the proposed method has to capture the change in the posterior distribution caused by all these variations. Only then can the formalism be employed for efficient forecasting of parameter constraints after a single initial full mock likelihood analysis needed to determine the Box-Cox parameters.

\subsection{Mock weak lensing survey}
\label{sec:survey}

Weak lensing surveys measure the shapes of millions of distant galaxy images which undergo tiny modifications when the light emitted by these galaxies is gravitationally lensed on its way to Earth. Correlating the shapes of pairs of galaxies, one can infer the statistical properties of the matter distribution projected along the line of sight, which in turn depends on the cosmological model. In addition, the weak lensing signal depends on the distances between observer, the structures acting as lenses, and the source galaxy, which provides information about the expansion history of the Universe. For details about gravitational lensing theory we refer the reader to \citet{bartelmann01}; for a recent review on weak lensing measurements see e.g. \citet{munshi08}.

While the majority of weak lensing studies use two-point correlation functions as the observable (see Section \ref{sec:corr}), predictions generally rely on Fourier space measures due to their direct connection to theory and their simple covariance properties. The power spectrum of the dimensionless projected mass density $\kappa$ reads \citep{kaiser92}
\eq{
\label{eq:limber}
C_\kappa(\ell) = \frac{9H_0^4 \Omega_{\rm m}^2}{4 c^4} \int^{\chi_{\rm hor}}_0 \dd \chi\; \frac{g^2(\chi)}{a^2(\chi)}\; P_\delta \br{\frac{\ell}{\chi},\chi}\;,
}
where $\ell$ denotes angular frequency, $H_0$ the Hubble constant, and $a$ the cosmological scale factor. The integral runs over comoving distance $\chi$ up to the horizon distance $\chi_{\rm hor}$. The power spectrum of the three-dimensional matter distribution is given by $P_\delta$, which depends on wavenumber $k=\ell/\chi$ and epoch, specified in terms of comoving distance. The geometrical contributions to equation (\ref{eq:limber}) are collected in the lensing efficiency
\eq{
\label{eq:lenseff}
g(\chi) = \int^{\chi_{\rm hor}}_{\chi} \dd \chi'\, n_{\rm g}\bb{z(\chi')}\, \br{1-\frac{\chi}{\chi'}}\;,
}
where $n_{\rm g}(z)$ denotes the normalised redshift distribution of galaxies in the survey.

We use $C_\kappa(\ell)$ as our weak lensing observable and evaluate it for 100 angular frequency bins, logarithmically spaced between $\ell_{\rm min}=10$ and $\ell_{\rm max}=10^4$. The fiducial cosmology used in our calculations is set to $\Omega_{\rm m}=0.25$, $\sigma_8=0.9$, the baryon density $\Omega_{\rm b}=0.05$, the power-law exponent of the initial matter power spectrum generated by inflation $n_{\rm s}=1.0$, and the Hubble parameter $h=0.7$, where $H_0 = h\, 100\,{\rm km/s/Mpc}$. Moreover the geometry of the Universe is assumed flat by default. To compute $P_\delta$, we employ the transfer function by \citet{eisenstein98} and apply the corrections due to non-linear evolution by \citet{PeacockDodds}.

The projected surface mass density is assumed to be Gaussian distributed, which implies that the covariance is given by \citep{joachimi08}
\eq{
\label{eq:covpower}
{\rm Cov}\bb{C_\kappa(\ell);C_\kappa(\ell')} = \delta_{\ell \ell'}\; \frac{4 \pi}{A_{\rm s}\, \ell\, \Delta \ell} \left( C_\kappa(\ell) + \frac{\sigma_\epsilon^2}{2\bar{n}_{\rm g}} \right)^2,
}
i.e. different angular frequencies are uncorrelated\footnote{Note that the assumption of Gaussianity is simplistic, in particular for high angular frequencies \citep[e.g.][]{kiessling11}, but still widely used for Fisher matrix forecasts (see \citealp{kiessling11b} though).}. Here, $\Delta \ell$ is the width of the angular frequency bin, and $A_{\rm s}=100\,{\rm deg}^2$ the survey size. The random orientations of the intrinsic shapes of source galaxies yield a shape noise contribution to equation (\ref{eq:covpower}), determined by the intrinsic ellipticity dispersion $\sigma_\epsilon=0.35$ and the total number density of galaxies on the sky $\bar{n}_{\rm g}=20\,{\rm arcmin}^{-2}$. We have implemented a redshift distribution of the form
\eq{
\label{eq:nofz}
n_{\rm g}(z) \propto z^2\; \exp \bc{-(z/z_0)^{1.5}}\;,
}
where the characteristic redshift scale $z_0$ is related to the median redshift via $z_0 \approx z_{\rm med}/1.4$. The survey is assumed to have a median redshift $z_{\rm med}=0.9$.

Following widespread practice, we make use of a Gaussian likelihood for $C_\kappa(\ell)$,
\eq{
\label{eq:cslikelihood}
L \propto \exp \bc{ - \frac{1}{2} \sum_{\ell=\ell_{\rm min}}^{\ell_{\rm max}} \frac{\bb{C_\kappa(\ell,\vek{p}_{\rm fid})-C_\kappa(\ell,\vek{p})}^2}{{\rm Cov}\bb{C_\kappa(\ell);C_\kappa(\ell)}}}\;,
}
where the power spectra obtained for the fiducial cosmology $\vek{p}_{\rm fid}$ serve as our mock data-vector. We assume flat priors and make sure that the likelihood peaks well inside the region of parameter space considered, so that the posterior is readily obtained from $L$ by renormalisation in parameter space.

\begin{figure*}   
\centering
\includegraphics[scale=.42,angle=270]{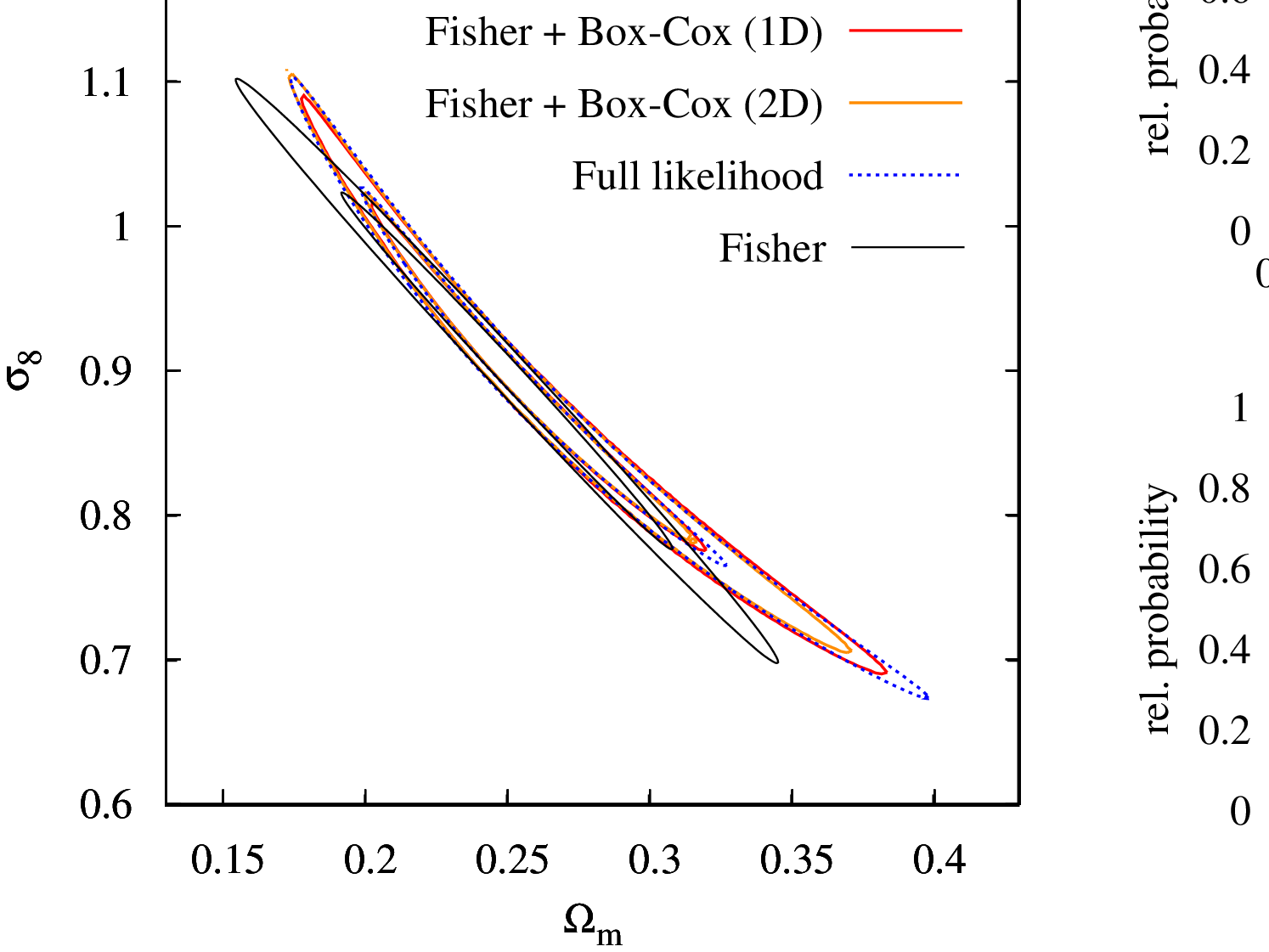}
\caption{\textit{Left panel}: $1\sigma$ and $2\sigma$ confidence levels in the $\Omega_{\rm m}-\sigma_8$ plane for the full likelihood analysis (blue dotted lines), the Box-Cox transformed posterior based on marginal distributions (red solid lines), and the Box-Cox-transformed posterior based on the two-dimensional distribution (orange solid lines). For comparison the results for a naive Fisher matrix computation are shown as black lines. \textit{Right panels}: Same as above, but for the marginalised distributions of $\Omega_{\rm m}$ (bottom) and $\sigma_8$ (top).}
\label{fig:parametertrafo}
\end{figure*} 

Again assuming Gaussianity, the corresponding Fisher matrix reads
\eq{
\label{eq:standardfisher}
F_{\mu\nu}^{\rm orig} = \sum_{\ell=\ell_{\rm min}}^{\ell_{\rm max}} \frac{\partial C_\kappa(\ell)}{\partial p_\mu}\; {\rm Cov}^{-1}\bb{C_\kappa(\ell);C_\kappa(\ell)}\; \frac{\partial C_\kappa(\ell)}{\partial p_\nu}\;,
}
where both the derivatives and the covariance are evaluated at $\vek{p}_{\rm fid}$. In writing equation (\ref{eq:standardfisher}) we have assumed that the covariance does not depend on cosmology; for the same reason we keep the covariance in equation (\ref{eq:cslikelihood}) fixed at its value for the fiducial set of cosmological parameters.

\subsection{Comparison of implementations}
\label{sec:comparison}

For most of the analysis we will only vary $\Omega_{\rm m}$ and $\sigma_8$ and keep all other cosmological parameters at their fiducial values. We compute the posterior on a grid in the $\Omega_{\rm m}-\sigma_8$ plane according to equation (\ref{eq:cslikelihood}) and also derive the marginal distributions for the two parameters. In Fig.$\,$\ref{fig:parametertrafo} we show confidence levels and marginal distributions for the likelihood analysis as well as for the standard Fisher matrix analysis using equation (\ref{eq:standardfisher}). While the marginal Fisher matrix errors on $\Omega_{\rm m}$ and $\sigma_8$ are still relatively close to the actual results, neither the tails in the marginal distributions, nor the banana-shaped form of the two-dimensional posterior and the extent of the confidence contours along the degeneracy can be reproduced by the standard Fisher matrix.

As a first step in the Box-Cox-Fisher formalism we determine the Box-Cox parameters from the full likelihood, using either the concentrated maximum likelihood from equation (\ref{eq:bclikelihood}) or the QQ-plot correlation coefficient from equation (\ref{eq:rqq}). The latter is restricted to one-dimensional distributions, i.e. in this case the marginal distributions of both $\Omega_{\rm m}$ and $\sigma_8$, whereas $L_{\rm max}$ is calculated for the individual marginal distributions as well as for the two-dimensional posterior. To obtain $L_{\rm max}$, a random sample of size $10^6$ is created from the respective distribution. The optimal values for $(\vek{\lambda},\vek{a})$ for which $L_{\rm max}$ or $r_{\rm QQ}$ attain a maximum are listed in Table \ref{tab:parametertrafo}.

\begin{table*}
\centering
\caption{Kullback-Leibler divergence $D_{\rm KL}$ between the posterior obtained from the full likelihood analysis and the posterior from the Box-Cox transformed Fisher matrices, using different implementations. Shown is $D_{\rm KL}$ for the distribution in the $\Omega_{\rm m}-\sigma_8$ plane in the second column, as well as for the marginalised distributions of $\Omega_{\rm m}$ and $\sigma_8$ in the third and fourth column. In the fifth to eighth column the optimum Box-Cox transformation parameters are listed for every implementation. Box-Cox parameters are either determined from the marginal distributions (1D) or the two-dimensional likelihood (2D). Results when using the different approaches to determining $\vek{p}_{\rm max}$ are also compared. For comparison results are also given for a standard Fisher matrix analysis.}
\begin{tabular}[t]{lrrrrrrr}
\hline\hline
analysis method & $D_{\rm KL}(\Omega_{\rm m})$ & $D_{\rm KL}(\sigma_8)$ & $D_{\rm KL}(\Omega_{\rm m},\sigma_8)$ & $\lambda(\Omega_{\rm m})$ & $a(\Omega_{\rm m})$ & $\lambda(\sigma_8)$ & $a(\sigma_8)$ \\
\hline
standard Fisher & 0.143 & 0.029 & 3.482 & - & - & - & - \\
Box-Cox 1D; $\vek{p}_{\rm max}$ from eq.$\,$(\ref{eq:pmaxbackward}) & 0.008 & 0.008 & 0.022 & -0.74 & 0.03 & 1.54 & 0.28 \\
Box-Cox 1D; $\vek{p}_{\rm max}$ from eq.$\,$(\ref{eq:pmaxforward}) & 0.008 & 0.007 & 0.043 & -0.74 & 0.03 & 1.54 & 0.28 \\
Box-Cox 1D; $\lambda,a$ via $L_{\rm max}$ & 0.010 & 0.008 & 0.040 & -0.09 & -0.08 & 3.73 & 4.00 \\
Box-Cox 2D; $\vek{p}_{\rm max}$ from eq.$\,$(\ref{eq:pmaxbackward}) & 0.013 & 0.013 & 0.017 & -0.03 & -0.03 & 0.87 & -0.46 \\
Box-Cox 2D; $\vek{p}_{\rm max}$ from eq.$\,$(\ref{eq:pmaxforward}) & 0.014 & 0.013 & 0.017 & -0.03 & -0.03 & 0.87 & -0.46  \\
\hline
\end{tabular}
\label{tab:parametertrafo}
\end{table*}

\begin{figure*}
\centering
\includegraphics[scale=.4,angle=270]{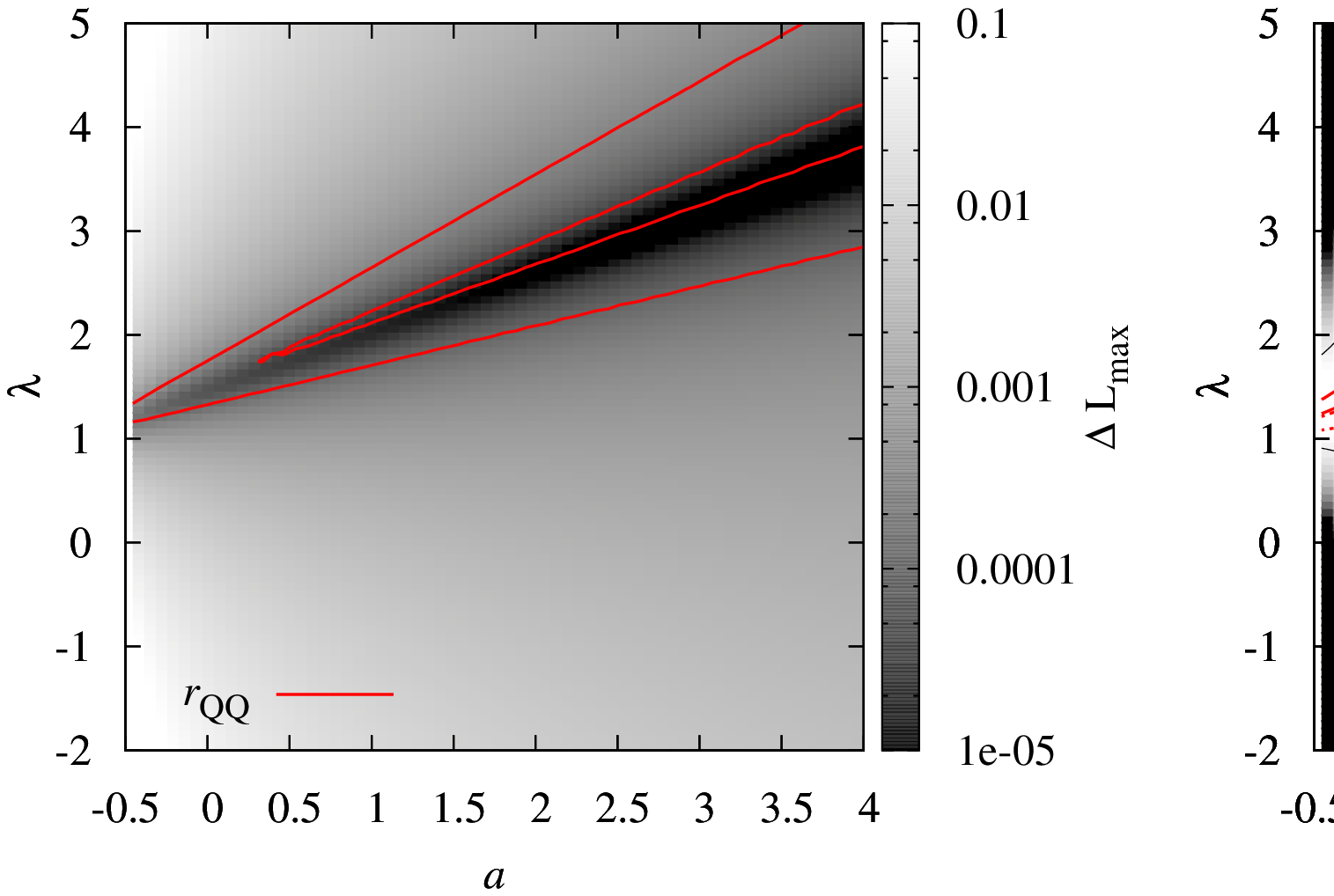}
\caption{\textit{Left panel}: Concentrated likelihood $L_{\rm max}$, see equation (\ref{eq:bclikelihood}), and QQ-plot correlation coefficient $r_{\rm QQ}$, see equation (\ref{eq:rqq}), for the marginalised distribution of $\sigma_8$, as a function of Box-Cox transformation parameters $\lambda$ and $a$. Red solid lines correspond to $r_{\rm QQ}$ and indicate a deviation of $10^{-5}$ and $10^{-4}$ from the maximum of 1. The relative deviation of $L_{\rm max}$ from its maximum is shown in grey shading, varying logarithmically between 0.1 (white) and $10^{-5}$ (black). \textit{Right panel}: Skewness and excess kurtosis of the transformed distribution as a function of $\lambda$ and $a$. Levels of constant skewness are shown in red, indicating values of 0.1, 0.01, -0.01, -0.1 from top to bottom. Contours for negative values are dotted. The kurtosis is shown in grey shading, varying linearly between 1 (black) and -0.1 (white). Levels of zero kurtosis are indicated by the black lines.}
\label{fig:bcplane_sigma8}
\end{figure*}

Working on one- or two-dimensional distributions, with $L_{\rm max}$ or $r_{\rm QQ}$ as statistic, results in largely different optimal values for the Box-Cox parameters. To gain further insight, we plot both statistics in the plane spanned by $\lambda$ and $a$ for the marginal distribution of $\sigma_8$ in Fig.$\,$\ref{fig:bcplane_sigma8}, left panel. Both $L_{\rm max}$ or $r_{\rm QQ}$ agree well in the region where they maximise. For a wide range in $(\lambda,a)$-space this maximum lies on a nearly perfect and almost linear degeneracy line. 

This degeneracy is mirrored in the shape of the Box-Cox transformed distribution, as can be seen in the right panel of Fig.$\,$\ref{fig:bcplane_sigma8}, where we show the skewness and excess kurtosis of the transformed distribution. The degeneracy in maximum $L_{\rm max}$ or $r_{\rm QQ}$ is closely matched by the minimum skewness with values close to zero. The kurtosis also features this degeneracy; however, it does not vanish, but instead obtains a shallow minimum at small negative values along the degeneracy line. Note that the skewness and kurtosis of the original distribution can be read off at $\lambda=1$. In this case contour lines are horizontal.

The mean and variance of the transformed distributions increase along the degeneracy line for larger values of $\lambda$ and $a$, so that the degeneracy can be broken by fixing either of the two lowest-order moments of the transformed distributions. However, since mean and variance are uncritical for our purposes, we leave them as free parameters and simply use the $(\lambda,a)$ combinations on the degeneracy line that our codes produce, the exact values hence determined by numerical effects and the maximisation algorithm used. See e.g. the values for $\lambda$ and $a$ in the third and fourth row of Table \ref{tab:parametertrafo} which lie in the region of maximum $L_{\rm max}$, $r_{\rm QQ}$ and minimum skewness. In the appendix we provide a toy model that illustrates basic properties of Box-Cox transformations including the degeneracy between $\lambda$ and $a$ discussed here.

With optimal values for $\lambda$ and $a$ at hand, we compute the transformed Fisher matrix as given in equation (\ref{eq:bcfisher2}) and subsequently the transformed posterior according to equation (\ref{eq:transformedposteriorfisher}). The resulting confidence contours and marginal distributions, with Box-Cox parameters obtained from the marginal distributions via $r_{\rm QQ}$ (1D) as well as from the full posterior via $L_{\rm max}$ (2D), are also shown in Fig.$\,$\ref{fig:parametertrafo}. Furthermore we provide a quantitative statement on how accurately the Box-Cox transformed posterior matches the actual one by calculating the Kullback-Leibler divergence $D_{\rm KL}$ as given by equation (\ref{eq:kullbackleibler}) between the two distributions in Table \ref{tab:parametertrafo}, again for both the two-dimensional and marginal cases.

Both visual and quantitative inspection demonstrate that the Box-Cox-Fisher formalism excellently reproduces the actual posterior, for all variants of the implementation considered. Compared to the standard Fisher results, the Box-Cox-Fisher formalism improves $D_{\rm KL}$ by a factor of 2 to 4 in the case of the marginal distribution of $\sigma_8$ and by at least an order of magnitude for the marginal distribution of $\Omega_{\rm m}$. The decrease in $D_{\rm KL}$ can mainly be ascribed to the accurate modelling of the non-Gaussian wings of the distributions, but partly also to the shift in the maximum of the marginal distributions away from the fiducial cosmology which the standard Fisher formalism cannot account for.

As the left-hand panel in Fig.$\,$\ref{fig:parametertrafo} suggests, the most blatant discrepancy between the standard and Box-Cox-Fisher analysis happens in the $\Omega_{\rm m}-\sigma_8$ plane, with about two orders of magnitude difference in $D_{\rm KL}$. The overall form of the posterior is represented accurately by the Box-Cox-Fisher contours; only the extent of the $2\sigma$ confidence levels reveals small residual deviations. As expected, if the Box-Cox parameters are derived from the marginal distributions, $D_{\rm KL}$ for the marginal distributions is smaller than for the 2D approach, and vice versa in the case of $D_{\rm KL}$ in the $\Omega_{\rm m}-\sigma_8$ plane. 

Note that we have also compared the algorithms given by equations (\ref{eq:pmaxforward}) and (\ref{eq:pmaxbackward}) to calculate $\vek{p}_{\rm max}$ in Table \ref{tab:parametertrafo}. Both perform equally well, but since equation (\ref{eq:pmaxbackward}) can be solved analytically for $\vek{p}_{\rm max}$, we will employ this version henceforth. Moreover we are going to apply the 2D approach, i.e. determining the Box-Cox parameters from the full posterior via $L_{\rm max}$, for the remainder of this paper.

\begin{figure}
\centering
\includegraphics[scale=.39,angle=270]{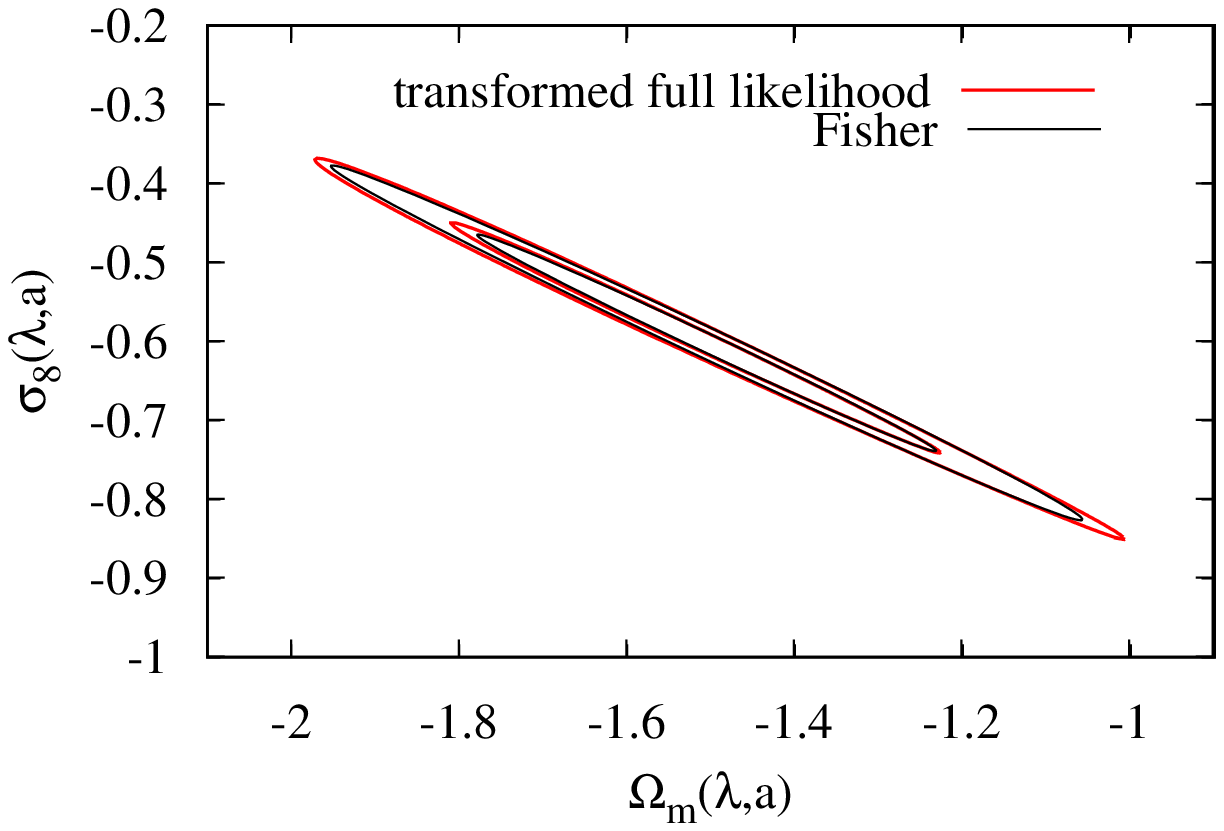}
\caption{$1\sigma$ and $2\sigma$ confidence levels in the plane of the Box-Cox transformed parameters. Red contours correspond to the transformed full likelihood, black contours originate from the transformed Fisher matrix given by equation (\ref{eq:bcfisher2}) and centred at $\bar{\vek{p}}_{\rm max}$. The posteriors are Gaussian to good approximation and agree well. The results shown were obtained for the case which is shown in Fig.$\,$\ref{fig:parametertrafo} as orange lines.}
\label{fig:parametertrafo2}
\end{figure} 

As seen in Fig.$\,$\ref{fig:bcplane_sigma8}, right panel, the optimal choice of Box-Cox parameters guarantees that the transformed distribution has vanishing skewness and low excess kurtosis, and hence can be assumed to be close to a multivariate Gaussian. Consulting equation (\ref{eq:transformedposteriorfisher}), this transformed distribution should additionally be described well by the transformed Fisher matrix, see equation (\ref{eq:bcfisher2}), with its peak at $\bar{\vek{p}}_{\rm max}$. This is illustrated in Fig.$\,$\ref{fig:parametertrafo2}, and indeed the transformed full posterior is closely matched by the transformed Fisher matrix contours. The slightly more extended confidence contours for the full likelihood might hint at a mildly platykurtic distribution which agrees with the small negative values of excess kurtosis along the degeneracy line in Fig.$\,$\ref{fig:bcplane_sigma8}. The ability of the Box-Cox transformations to change the posterior into a multivariate Gaussian opens up a range of potential applications, as we will discuss further in Section \ref{sec:conclusions}.

\subsection{Varying cosmology and survey parameters}
\label{sec:scaling}

\begin{table*}
\centering
\caption{Kullback-Leibler divergence $D_{\rm KL}$ between the posterior obtained from the full likelihood analysis and the posterior from the Box-Cox transformed Fisher matrices, varying different survey or cosmological parameters as indicated in the first column. Shown is $D_{\rm KL}$ for the distribution in the $\Omega_{\rm m}-\sigma_8$ plane in the second column, as well as the marginalised distributions for $\Omega_{\rm m}$ and $\sigma_8$ in the third and fourth column.}
\begin{tabular}[t]{lrrr}
\hline\hline
parameters changed & $D_{\rm KL}(\Omega_{\rm m})$ & $D_{\rm KL}(\sigma_8)$ & $D_{\rm KL}(\Omega_{\rm m},\sigma_8)$\\
\hline
fiducial parameters & 0.013 & 0.013 & 0.017 \\
$\Omega_{\rm m}: 0.25 \rightarrow 0.225$ & 0.008 & 0.008 & 0.025 \\
$z_{\rm med}: 0.9 \rightarrow 1.0$; $n_{\rm g}: 20 \rightarrow 37.4\,{\rm arcmin}^{-2}$ & 0.005 & 0.005 & 0.019 \\
$\ell_{\rm max}: 10000 \rightarrow 8700$ & 0.015 & 0.015 & 0.019 \\
$A_{\rm s}: 100 \rightarrow 110\,{\rm deg}^2$ & 0.013 & 0.012 & 0.016 \\
\hline
\end{tabular}
\label{tab:parametertrafo2}
\end{table*}

We expect the Box-Cox-Fisher formalism to be particularly useful in an advanced planning stage of an experiment when e.g. the capabilities of breaking model parameter degeneracies come into focus. By then the survey parameters and the analysis strategies should not change radically anymore, but only in relatively small steps and only a few parameters at a time. If that holds true, the general form of the posterior is only moderately modified under these changes, so that one can continue to use the optimal Box-Cox parameters determined from the initial full likelihood analysis.

The Box-Cox-Fisher analysis is repeated for several survey configurations that each differ in one or two parameters from the fiducial survey by about $10\,\%$. These changes are accounted for in the Fisher matrix, but we retain the values of the Box-Cox parameters determined for the fiducial survey. A full likelihood analysis is computed as well for every configuration, but solely for the purpose of assessing the accuracy of the forecast.

We modify the fiducial cosmology by lowering $\Omega_{\rm m}$ by $10\,\%$. A slightly deeper survey is analysed, increasing $z_{\rm med}$ to 1, which also increases the number density of galaxies and consequently reduces the noise contribution. Applying the scaling found by \citet{amara07}, the deeper survey has $\bar{n}_{\rm g}=37.4\,{\rm arcmin}^{-2}$. Moreover we consider the case of discarding the highest angular frequency bins in the analysis, reducing $\ell_{\rm max}$ to 8700. Finally, we increase the survey size by $10\,\%$.

Analogously to the foregoing section, we employ the Kullback-Leibler divergence to compare the Box-Cox-Fisher result with the posterior from the full likelihood analysis. As is evident from Table \ref{tab:parametertrafo2}, $D_{\rm KL}$ for the marginal distributions and the posterior in the $\Omega_{\rm m}-\sigma_8$ plane remains constant to good approximation in all cases.

\subsection{Varying statistic and posterior dimension}
\label{sec:corr}

One of the most likely modifications in mock weak lensing analyses is a change in the statistic used as the observable. We switch to the frequently employed correlation function $\xi_+$ which is related to the power spectrum via \citep{SvWM02}
\eq{
\label{eq:xiplus}
\xi_+(\theta) = \int^\infty_0 \frac{\dd \ell\, \ell}{2 \pi}\; J_0(\ell \theta)\; C_\kappa(\ell)\;,
}
where $J_0$ is the Bessel function of the first kind of order 0. The covariance of the correlation function can directly be determined from equation (\ref{eq:covpower}), as detailed in \citet{joachimi08}. We intend to roughly use the same angular scales as in the power spectrum analysis and thus consider the range $1\,{\rm arcmin}<\theta<5\,{\rm deg}$, divided into 50 logarithmically spaced bins. Note that this range of angular scales does not ensure a similar information content because the angular separation bins are strongly correlated.

\begin{figure*}
\centering
\includegraphics[scale=.42,angle=270]{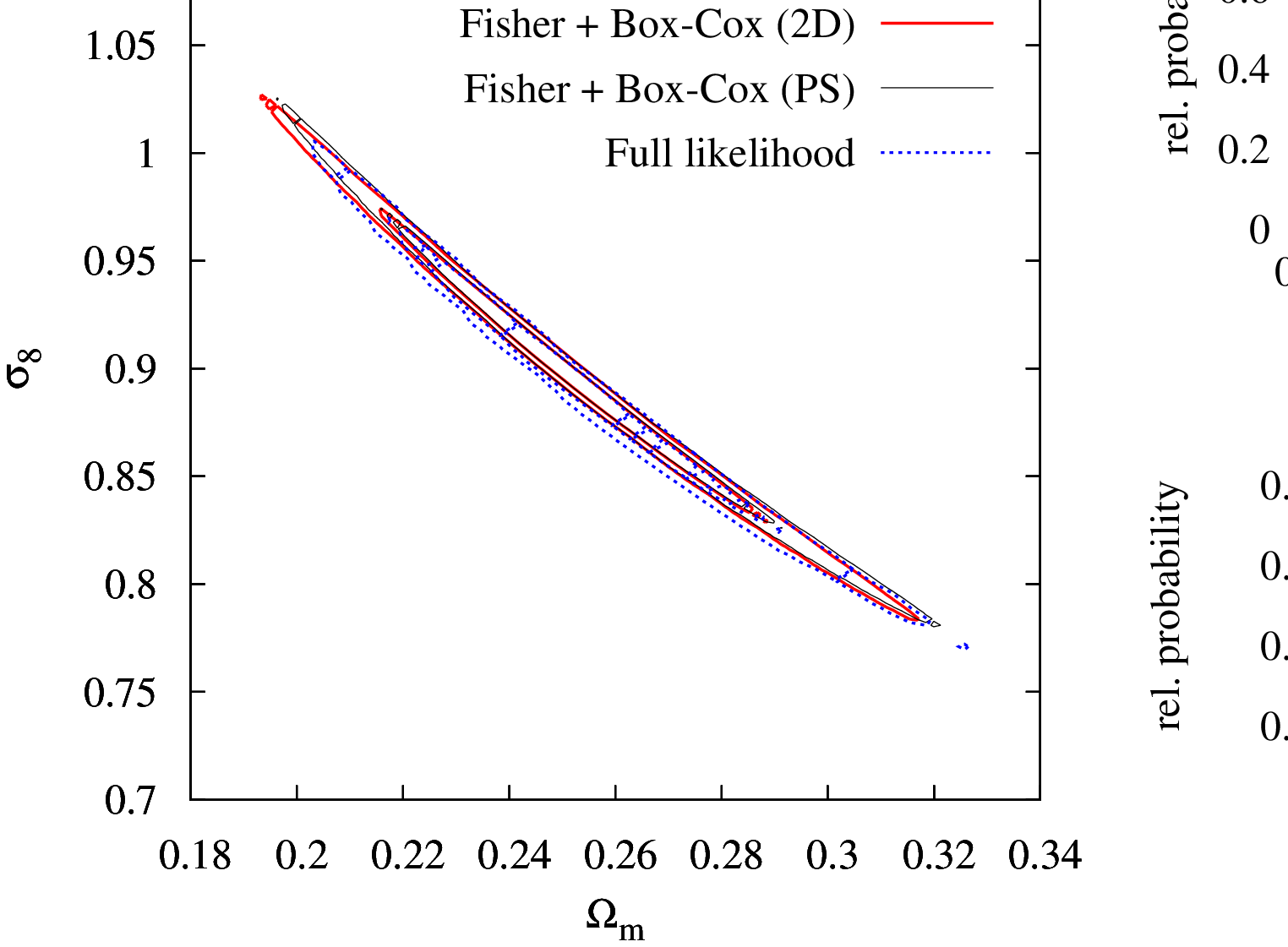}
\caption{\textit{Left panel}: $1\sigma$ and $2\sigma$ confidence levels in the $\Omega_{\rm m}-\sigma_8$ plane for the full likelihood analysis (blue dotted lines) and the Box-Cox transformed posterior (red solid lines), using the shear correlation function $\xi_+$ as statistic. The optimal Box-Cox parameters obtained from the power spectrum analysis also yield good results in this case, as indicated by the black solid lines. \textit{Right panels}: Same as above, but for the marginalised distributions of $\Omega_{\rm m}$ (bottom) and $\sigma_8$ (top).}
\label{fig:parametertrafo_corr_2d}
\end{figure*} 

Moreover we now use Population Monte-Carlo sampling with CosmoPMC\footnote{\texttt{http://www2.iap.fr/users/kilbinge/CosmoPMC/}} \citep{cappe08,wraith09} to create a random sample of size $10^5$ from the full posterior in order to determine optimal Box-Cox parameters via equation (\ref{eq:bclikelihood}). The results are presented in Fig.$\,$\ref{fig:parametertrafo_corr_2d}, finding again excellent agreement between Box-Cox-Fisher results and full posterior. If the optimal Box-Cox parameters that were obtained for the power spectrum analysis in Section \ref{sec:comparison} are used instead, one arrives at constraints of similar quality. Therefore the Box-Cox-Fisher formalism should also be robust with respect to a change in the weak lensing statistic employed in the Fisher matrix.

As a final test for the practical applicability of the novel forecasting method, we have to verify that it is accurate for a higher-dimensional posterior. Hence we drop the assumption of a spatially flat Universe and vary the density parameter of dark energy, $\Omega_\Lambda$, as well as $n_{\rm s}$ in addition to $\Omega_{\rm m}$ and $\sigma_8$. We perform the analysis as in the foregoing case, using again CosmoPMC to create about $10^6$ random samples of the four-dimensional posterior to determine in total 8 Box-Cox parameters.

\begin{figure*}
\centering
\includegraphics[scale=.38,angle=270]{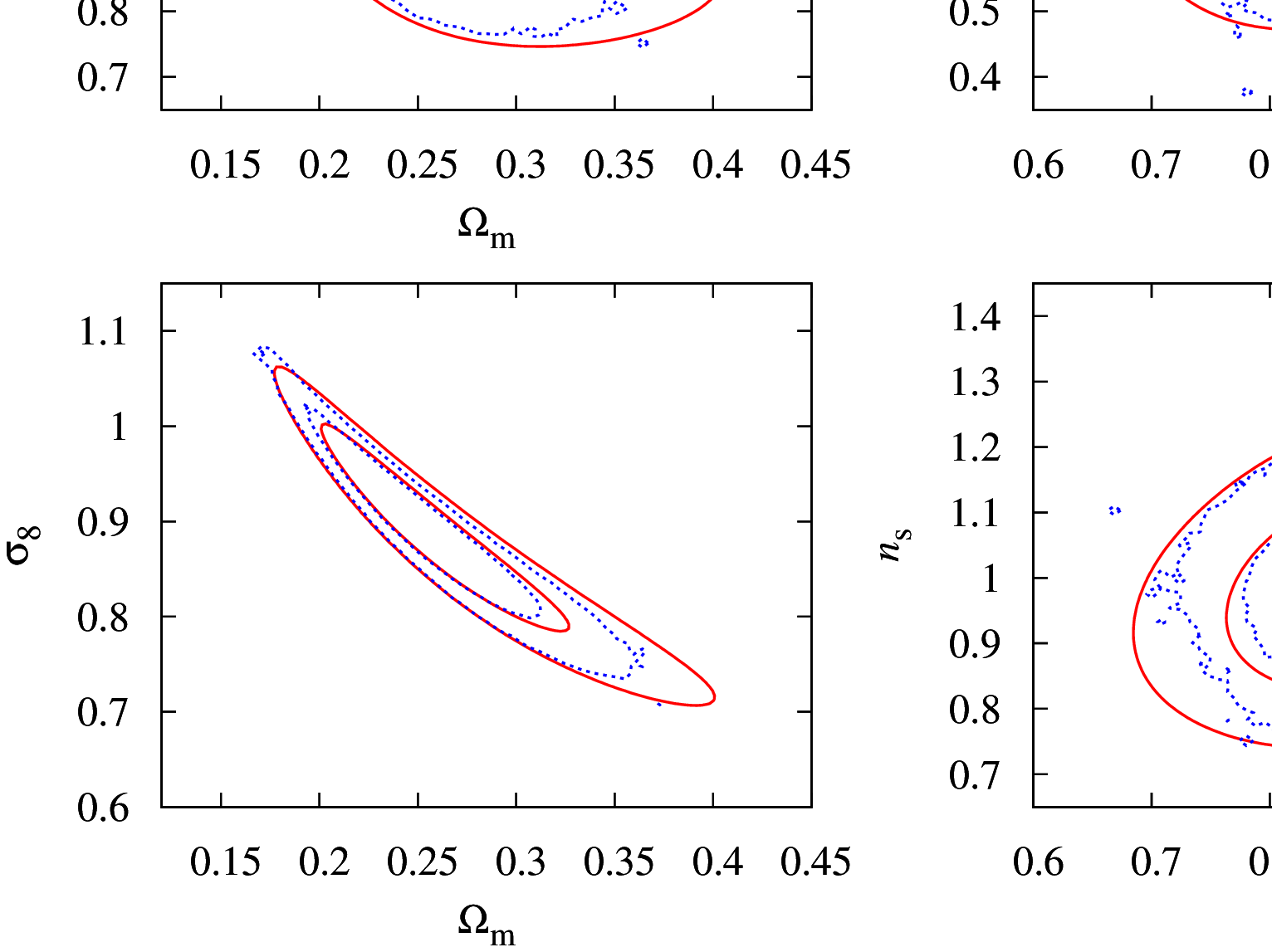}
\caption{$1\sigma$ and $2\sigma$ confidence levels for the full likelihood analysis (blue dotted lines) and the Box-Cox transformed posterior (red solid lines) for all two-dimensional marginalised distributions in the four-dimensional parameter space $\bc{\Omega_{\rm m},\sigma_8,\Omega_\Lambda,n_{\rm s}}$. Note that the Fisher matrix as computed by CosmoPMC has been employed in the Box-Cox analysis.}
\label{fig:parametertrafo_corr_4d}
\end{figure*} 

The confidence contours of the marginalised posterior distributions for all possible pairs of cosmological parameters are shown in Fig.$\,$\ref{fig:parametertrafo_corr_4d}. The Box-Cox-Fisher formalism yields contours that are able to adopt arbitrary shapes and represent the four-dimensional posterior accurately, including the non-linear degeneracies in the $\Omega_{\rm m}-\sigma_8$ and $\Omega_\Lambda-n_{\rm s}$ planes. The only significant discrepancies between the Fisher-based contours and the confidence levels derived from the Monte-Carlo sample appears in regions where the posterior declines slowly, e.g. for large $\Omega_{\rm m}$ or small $\sigma_8$. The frayed contour lines indicate that these regions are still sparsely sampled by CosmoPMC. This could imply that the Monte-Carlo sample is not suited to allow for a determination of optimal Box-Cox parameters which lead to an accurate posterior shape in these regions. Alternatively, the Box-Cox-Fisher formalism might well be robust enough to produce a precise representation of the posterior also where it is shallow, so that the difference in contour lines would be caused by the insufficient Monte-Carlo sampling in that regime.

As a further example for the reliability of the Box-Cox-Fisher formalism, we initially observed a slight tilt of the Box-Cox-Fisher confidence contours against those from the Monte-Carlo analysis, particularly in the $\Omega_\Lambda-n_{\rm s}$ plane, which could be traced back to a small difference in the correlation functions computed by CosmoPMC and the authors' code. The latter was used to produce the fiducial $\xi_+$ which served as the mock datavector input to CosmoPMC. The small discrepancy in $\xi_+$ leads to a small shift in the maximum likelihood point as determined by CosmoPMC away from the fiducial cosmolgy, as well as slightly different derivatives of $\xi_+$ with respect to cosmological parameters. Consequently, the authors' code and CosmoPMC produce moderately discrepant Fisher matrices. Using the latter in the Box-Cox-Fisher analysis instead results in the excellent agreement shown in Fig.$\,$\ref{fig:parametertrafo_corr_4d}.

\section{An application: Breaking degeneracies in the $\Omega_{\rm m}-\sigma_8$ plane}
\label{sec:application}

\begin{figure*}
\centering
\includegraphics[scale=.42,angle=270]{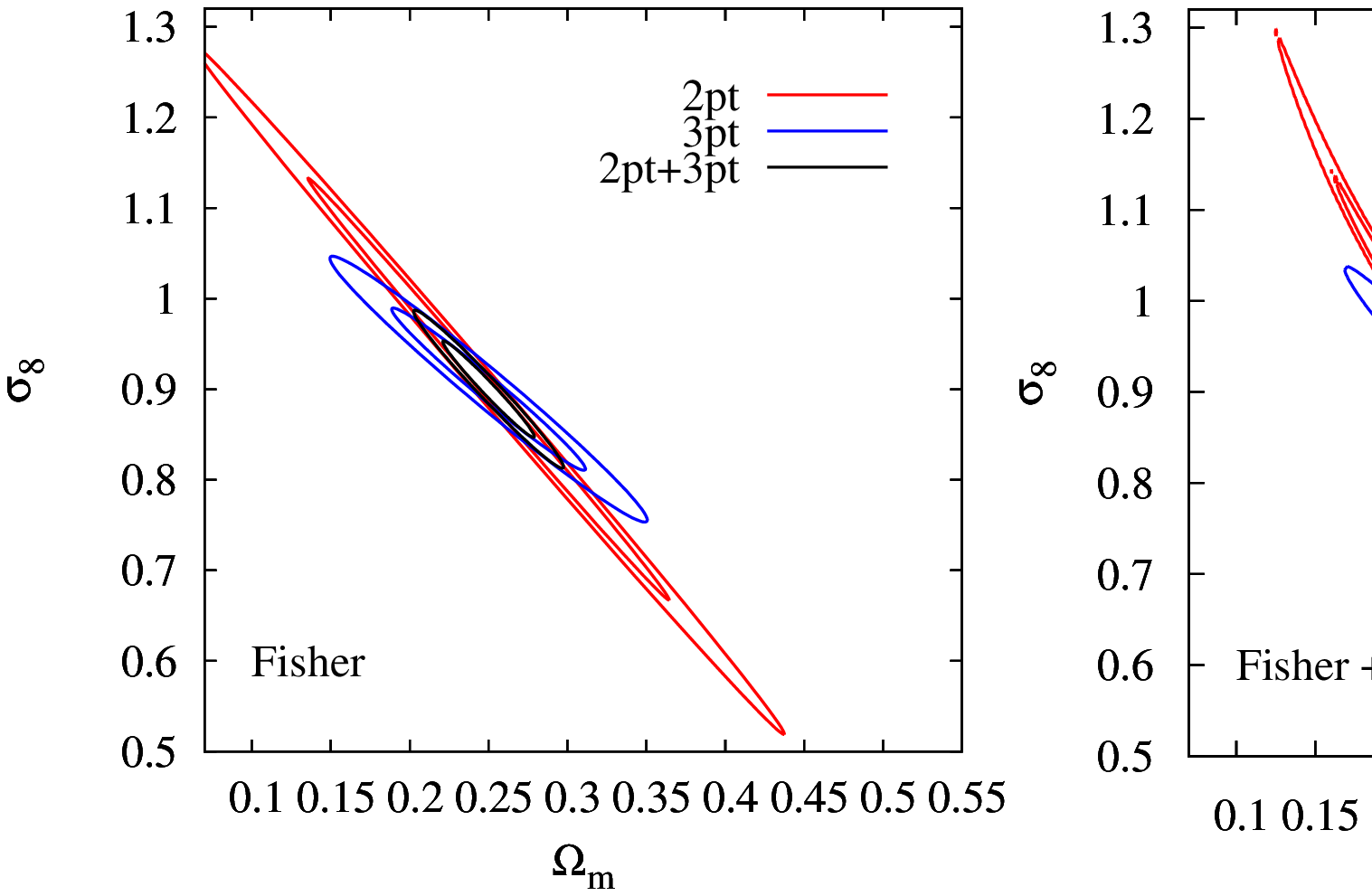}
\caption{Combined power spectrum and bispectrum constraints on $\Omega_{\rm m}$ and $\sigma_8$. \textit{Left panel}: $1\sigma$ and $2\sigma$ confidence levels for a standard Fisher matrix analysis of two-point weak lensing statistics (red lines), three-point statistics (blue lines), as well as two- and three-point statistics combined (black lines). \textit{Right panel}: Same as above, but for constraints resulting from the Box-Cox-Fisher analysis.}
\label{fig:bispectrum_degeneracy}
\end{figure*} 

The Box-Cox-Fisher formalism is applicable to a wide range of problems. For illustrational purposes we provide in the following a toy example which is again built around a mock weak lensing survey. Future experiments which are currently in the planning stages will not be restricted to measuring two-point statistics like $C_\kappa(\ell)$, but make use of higher-order correlations of galaxy shapes. Three-point statistics such as the bispectrum $B_\kappa(\ell_1,\ell_2,\ell_3)$ have been demonstrated to potentially tighten cosmological parameter constraints considerably, e.g. by breaking degeneracies in the $\Omega_{\rm m}-\sigma_8$ plane \citep{berge10}. Since these conclusions rely entirely on standard Fisher analyses, we set out to investigate whether the breaking of $\Omega_{\rm m}-\sigma_8$ degeneracies is affected by the actual, non-elliptical shapes of confidence levels for both two- and three-point weak lensing statistics.

We treat $B_\kappa(\ell_1,\ell_2,\ell_3)$ as our observable three-point statistic and calculate it via \citep[e.g.][]{takada04}
\eqa{
\label{eq:bispectrum}
B_\kappa(\ell_1,\ell_2,\ell_3) &=& \frac{27H_0^6 \Omega_{\rm m}^3}{8 c^6} \int^{\chi_{\rm hor}}_0 \dd \chi\; \frac{g^3(\chi)}{\chi\; a^3(\chi)}\\ \nn
&& \times\; B_\delta \br{\frac{\ell_1}{\chi},\frac{\ell_2}{\chi},\frac{\ell_3}{\chi},\chi}\;,
}
where $B_\delta$ denotes the matter bispectrum. We employ perturbation theory \citep{fry84} to compute $B_\delta$ from the matter power spectrum, applying the corrections due to non-linear structure evolution given in \citet{scoccimarro01}. We employ the bispectrum covariance according to \citet{joachimi09b}, using only the lowest-order term that is given in terms of power spectra. Noting that the bispectrum is only non-zero if its three arguments can form the sides of a triangle (see e.g. \citealp{joachimi09b} for details), we assemble the datavector out of all such combinations, where $\ell_1,\ell_2,\ell_3$ can have 20 logarithmically spaced values between 10 and 1000.

Performing a full mock likelihood analysis for the bispectrum is computationally costly, even if only two cosmological parameters are varied. As a bi-product from another project, we have bispectrum computations on a $20 \times 20$ grid in the $\Omega_{\rm m}-\sigma_8$ plane at our disposal, albeit for a different cosmology than the fiducial survey outlined in Section \ref{sec:survey}. The grid was created for a fiducial cosmology with deviating parameters $\Omega_{\rm b}=0.045$, $h=0.71$, and $n_{\rm s}=0.963$. Furthermore the non-linear correction for the matter power spectrum by \citet{smith03} was used. The bispectra were obtained for a single source redshift, i.e. the redshift distribution in equation (\ref{eq:nofz}) is replaced by a Dirac delta-distribution peaking at $z_{\rm s}=1$. Finally, the angular frequency binning is slightly different, with 18 bins between 10 and 1500.

We make use of the scaling properties of the Box-Cox-Fisher formalism and determine optimal Box-Cox parameters for the bispectrum from a mock likelihood analysis based on the gridded bispectra. The changes in cosmology are not more than $10\,\%$. Besides, they only affect parameters that are kept fixed in this analysis, and that weak lensing is less sensitive to than $\Omega_{\rm m}$ and $\sigma_8$. The single source redshift, $z_{\rm s}$, is similar to the median redshift of the fiducial survey, so that the lensing efficiency should change only mildly, see equation (\ref{eq:lenseff}). Likewise, the different non-linear corrections and angular frequency coverage should not alter the shape of the posterior in the $\Omega_{\rm m}-\sigma_8$ plane significantly.

In the power spectrum analysis we adopt the Box-Cox parameters determined for the fiducial survey. We also use the fiducial survey parameters, except for $\ell_{\rm max}$ which is reduced to 3000. The low angular frequency cut-offs for both two- and three-point statistics are meant to exclude the deeply non-linear clustering regime and thus improve the simplistic approximations for the covariances, particularly for the bispectrum. Additionally, we combine the constraints from two- and three-point statistics by simply adding Fisher matrices or multiplying posteriors, respectively, i.e. we assume that power spectra and bispectra are uncorrelated (which, again, is simplistic but common practice, e.g. \citealp{berge10}). 

\begin{table}
\centering
\caption{Marginalised $2\sigma$ constraints on $\Omega_{\rm m}$ and $\sigma_8$ resulting from the standard Fisher and Box-Cox-Fisher analyses of two-point, three-point, and combined two- and three-point weak lensing statistics.}
\begin{tabular}[t]{llrr}
\hline\hline
method & statistics & $\Omega_{\rm m}$ & $\sigma_8$ \\
\hline
                 & 2pt       & $0.25^{+0.11}_{-0.11}$ & $0.90^{+0.23}_{-0.23}$ \\
Fisher           & 3pt       & $0.25^{+0.06}_{-0.06}$ & $0.90^{+0.09}_{-0.09}$ \\
                 & 2pt + 3pt & $0.25^{+0.03}_{-0.03}$ & $0.90^{+0.05}_{-0.05}$ \\
\hline
                 & 2pt       & $0.25^{+0.22}_{-0.11}$ & $0.85^{+0.31}_{-0.27}$ \\
Fisher + Box-Cox & 3pt       & $0.25^{+0.10}_{-0.07}$ & $0.89^{+0.11}_{-0.12}$ \\
                 & 2pt + 3pt & $0.25^{+0.05}_{-0.03}$ & $0.90^{+0.06}_{-0.08}$ \\
\hline
\end{tabular}
\label{tab:bispectrum_degeneracy}
\end{table}

In Fig.$\,$\ref{fig:bispectrum_degeneracy} we contrast the parameter constraints in the $\Omega_{\rm m}-\sigma_8$ plane from the standard Fisher matrix and the Box-Cox-Fisher analysis for two-point statistics, three-point statistics, and both data sets combined. The posterior for the bispectrum constraints alone also features the characteristic $\Omega_{\rm m}-\sigma_8$ degeneracy, albeit with a tilted degeneracy line, a property that is also captured by the standard Fisher matrix \citep[see also][]{takada04,berge10}. Since the intersection of the contours is at a sufficiently large angle, the joint constraints in the Box-Cox-Fisher case produce fairly elliptical confidence contours which are of similar size as those resulting from the standard Fisher matrix.

The marginalised constraints on $\Omega_{\rm m}$ and $\sigma_8$ presented in Table \ref{tab:bispectrum_degeneracy} allow for a more quantitative evaluation. Taking into account the accurate shape of the posterior generally increases the $2\sigma$ confidence interval substantially. This increase is stronger the greater the deviation of the posterior from a Gaussian shape, see e.g. the increase by $50\,\%$ for $\Omega_{\rm m}$ in the power spectrum analysis. In the case of the joint two- and three-point constraints, the absolute change in errors is smaller, but still the $2\sigma$ confidence interval grows by about $40\,\%$ ($30\,\%$) for $\sigma_8$ ($\Omega_{\rm m}$).

\section{Conclusions}
\label{sec:conclusions}

In this work we introduced a novel method to compute precise predictions for statistical constraints on model parameters from future experiments. By combining two generic statistical tools -- the Fisher matrix and Box-Cox transformations, we were able to drop the assumption of Gaussianity in parameter space. Applying Box-Cox transformations to model parameters, one arrives at approximately multivariate Gaussian shapes of the posterior. In this transformed space the Fisher matrix can be computed without suffering from the usual limits of the Gaussian assumption. An inverse Box-Cox transformation of the Fisher matrix results then yields realistic posterior distributions in the original parameter space.

We derived the formalism of the combined Fisher and Box-Cox analysis and detailed different approaches to determining the parameters of the Box-Cox transformation from an inital likelihood analysis. Utilising a mock weak lensing survey, we verified the accuracy of the Box-Cox-Fisher formalism and demonstrated that it robustly accounts for changes in various survey parameters and analysis steps. We expect the method to be particularly useful in the advanced planning stages of upcoming experiments and surveys, e.g. to fine-tune the design with repect to the anticipated parameter constraints, or to quantify the breaking of model parameter degeneracies when combining data sets.

A practical implementation of the Box-Cox-Fisher formalism can look as follows:
\begin{enumerate}
\item Obtain information about the full likelihood for a fiducial experiment, for instance from a gridded likelihood in parameter space or via Monte-Carlo sampling.
\item Determine the optimal Box-Cox parameters using the statistics $L_{\rm max}$ or $r_{\rm QQ}$, see equations (\ref{eq:bclikelihood}) and (\ref{eq:rqq}).
\item Calculate the standard Fisher matrix for the exact experimental setup one is interested in.
\item Compute the posterior via equations (\ref{eq:transformedposteriorfisher}), (\ref{eq:bcfisher2}), and (\ref{eq:pmaxbackward}).
\end{enumerate}
The last two steps can be repeated as required for arbitrary values of experimental parameters, as long as these changes do not alter the shape of the posterior too strongly from the initial likelihood analysis. This might for instance happen if several experimental parameters are varied substantially at the same time. Unfortunately, adding new model parameters to the analysis also potentially modifies the posterior significantly, depending on the correlation of this new parameter with the existing ones. Therefore we consider it unlikely that Box-Cox parameters can be determined to sufficent accuracy from low-dimensional sub-spaces of the posterior distribution.

The price to pay for the dramatically more realistic posteriors and confidence regions compare to the standard Fisher analysis is the need for an initial determination of the Box-Cox parameters which requires detailed information about the full posterior distribution. For a realistic number of model parameters the sampling or gridded evaluation of the likelihood is a computationally expensive step. However, complex experiments demand in practice hundreds of forecast calculations, so that switching to a Box-Cox-Fisher prediction after an initial full mock likelihood analysis is still largely beneficial in terms of computational time. Note that the extra calculations required for the new method add only marginally to the time the corresponding standard Fisher matrix computation takes.

We illustrated a potential application of the Box-Cox-Fisher formalism, investigating the effects of precise posterior modelling on the joint constraints by weak lensing two- and three-point statistics in the $\Omega_{\rm m}-\sigma_8$ plane. We find that while the shapes of confidence contours for the individual constraints from power spectra and bispectra change in a pronounced way from the standard Fisher results, the joint posterior is compact and close to the Gaussian form predicted by the standard Fisher matrix, hence confirming in the simple case we considered that three-point statistics can indeed break the $\Omega_{\rm m}-\sigma_8$ degeneracy to a large extent. However, marginal errors on the cosmological parameters increase substantially by up to $50\,\%$ when using the Box-Cox-Fisher analysis instead of standard Fisher matrix forecasts, which certainly needs to be taken into account for predictions of precision measurements.

Generally, the more compact a posterior is, the more it looks Gaussian. Consequently, the local representation around the maximum provided by the Fisher matrix provides a good description of the complete posterior shape in that case. It should be noted that, in order to provide a challenging benchmark to test our method, and to facilitate the covariance calculations, we deliberately designed our exemplary weak lensing survey to yield only weak cosmological constraints. Future weak lensing surveys will perform much better, thereby rendering the Gaussian approximation in parameter space more appropriate for predictions. Furthermore, combining different cosmological probes yielding orthogonal constraints helps breaking parameter degeneracies and thus also renders the posterior more compact, so that the standard Gaussian Fisher matrix should perform comparatively well in such joint parameter analyses.

Yet, experiments will always be faced with complex posterior distributions. For instance upcoming large-area weak lensing surveys will be used to test modifications of gravity. A popular parametrisation of deviations from General Relativity introduces the gravitational slip and a modification of Newton's constant, where the two parameters are perfectly degenerate and non-linearly related for weak lensing data alone \citep[see e.g.][]{daniel10b}. Comparing the Fisher matrix forecasts for these parameters in \citet{guzik09} with the likelihood analyses in \citet{daniel10b} and \citet{song10}, it is evident that the optimisation of the survey design for modified gravity measurements will need to go beyond the standard Fisher matrix approach.

For future developments of the Box-Cox-Fisher formalism it will prove fruitful to continue the data analysis in the Box-Cox transformed parameter space, and not transform back to the physically motivated model parameters, as done in this work. Then one can fully exploit the Gaussian form of the posterior and apply the whole arsenal of statistical tools that become accurate, or usable in the first place, on Gaussian distributions. One such application, which will be dealt with in a forthcoming publication, is the subsequent decorrelation of the Box-Cox transformed model parameters, which may open up the possibility to define statistically independent variables.

Many steps in statistical data analysis are simplified or improve in accuracy when working with Gaussian distributions, so that one can potentially benefit from Box-Cox transformations in a wide range of problems. For example, \citet{taylor10} have developed an analytical marginalisation technique that works on Gaussian subspaces of the posterior. Using Box-Cox transformations, one can transform sub-spaces of or the complete posterior to a multivariate Gaussian, and moreover assess the non-Gaussianity of a given model parameter to verify whether a transformation is required. 

Monte-Carlo Markov Chain (MCMC) methods sample a posterior considerably more efficiently if the distribution is compact and does not feature low-probability tails along degeneracy directions. As an example consider the cosmic microwave background (CMB) likelihood analysis by \citet{tegmark04} who employed the \lq natural\rq\ parametrisation suggested by \citet{kosowsky02}, followed by the diagonalisation of the parameter covariance matrix. Analogously, one could use an intial coarse MCMC sample to determine Box-Cox transformations that render the posterior approximately Gaussian. After an additional decorrelation of parameter space the detailed MCMC analysis could be run with high efficiency on a set of model parameters which are statistically independent and Gaussian distributed to good accuracy. This ansatz is applicable to any kind of likelihood analysis and does not require the existence of a physically motivated set of natural parameters, as in the case of the CMB.

Note furthermore that logarithmic transformations, which constitute a special case of Box-Cox transformations, of the large-scale matter distribution or the weak lensing convergence have recently been shown to enhance the information content of two-point statistics \citep{neyrinck09,seo11}. The potential of the more general Box-Cox transformations in this case is currently under investigation.

\section*{Acknowledgments}

We would like to thank Martin Hendry and Peter Schneider for fruitful discussions, and our referee for an encouraging report. Moreover we acknowledge the help of Martin Kilbinger and Eric Tittley in setting up CosmoPMC. We are grateful to Martin Kilbinger for making CosmoPMC publicly available. BJ acknowledges support by the European DUEL network, project MRTN-CT-2006-036133, and a UK Space Agency Euclid grant.

\bibliographystyle{mn2e}

\appendix

\section{Illustration of Box-Cox transformations}

In the following we will provide a toy model that illustrates the principle of Box-Cox transformations, reproducing in particular the nearly linear degeneracy between the Box-Cox parameters $\br{\lambda,a}$ encountered in Section \ref{sec:comparison}.

\begin{figure}
\centering
\includegraphics[scale=.38,angle=270]{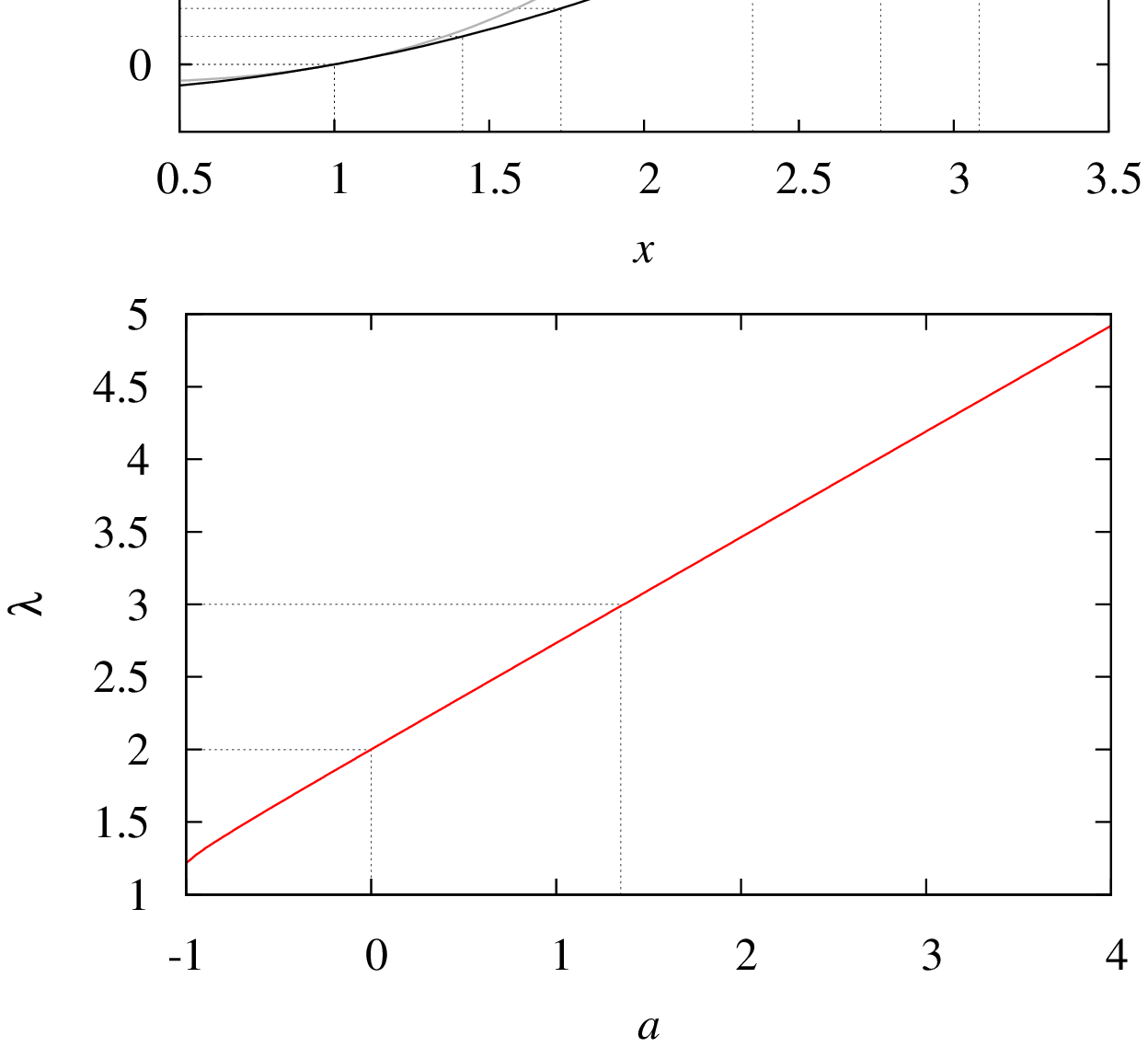}
\caption{\textit{Upper panel}: Illustration of Box-Cox transformations. Shown are the transformations for $\lambda=2$ in black and for $\lambda=3$ in grey. The set of black lines in the bottom left corner indicates the transformation of the data set $\bc{1,\sqrt{2},\sqrt{3}}$ with $\br{\lambda=2;a=0}$, the other set the transformation of the same data with $\br{\lambda=3;a=1.35}$. Note that in both cases the transformed data values are equidistant. \textit{Lower panel}: Pairs of Box-Cox parameters $\br{\lambda,a}$ for which the transformed data has zero skewness. The black lines correspond to the two cases shown in the upper panel. Note that the relation between $\lambda$ and $a$ is linear over a wide range.}
\label{fig:modeldegeneracy}
\end{figure} 

Suppose one wanted to transform the data set $\bc{1,\sqrt{2},\sqrt{3}}$, using Box-Cox transformations, such that the transformed data have vanishing skewness. An obvious choice in this case is to simply square the data, i.e. apply a Box-Cox transformation with $\br{\lambda=2;a=0}$. As is demonstrated in Fig.$\,$\ref{fig:modeldegeneracy}, this indeed renders the transformed data values equidistant and hence unskewed.

More generally, the requirement of zero skewness implies in the case of three elements $\bc{x_1,x_2,x_3}$ in the datavector the condition $\bar{x}_3-\bar{x}_2=\bar{x}_2-\bar{x}_1$, assuming the transformation does not change the ordering of the data. As before, the bar denotes the transformed data values. Inserting the definition of the Box-Cox transformation as given in equation (\ref{eq:bctrafo}), one obtains
\eq{
\label{eq:toymodelcondition}
\br{x_1+a}^\lambda + \br{x_3+a}^\lambda = 2 \br{x_2+a}^\lambda\;.
}
The solutions of this equation for the data set $\bc{x_1=1;\,x_2=\sqrt{2};\,x_3=\sqrt{3}}$ are plotted in the lower panel of Fig.$\,$\ref{fig:modeldegeneracy}, revealing a linear relation between $\lambda$ and $a$, except in the regime $a \leq 0$.

From this relation one can read off more combinations of $\br{\lambda,a}$ that should fulfil equation (\ref{eq:toymodelcondition}), e.g. $\br{\lambda=3;a=1.35}$. As is shown in the top panel of the figure, shifting the data to larger values by 1.35 and then taking the third power again results in an unskewed transformed distribution, albeit with a larger mean and variance than for the case $\br{\lambda=2;a=0}$. 

In summary, this toy model reproduces the findings from Fig.$\,$\ref{fig:bcplane_sigma8}. Optimally removing the skewness of a data set (a task which also seems to govern the Gaussianisation of posteriors considered in this work) determines the Box-Cox parameters up to a perfect degeneracy which is very close to linear over a wide range of the $\br{\lambda,a}$ plane. Along the degeneracy line the mean and variance of the transformed distribution vary. Interestingly, even in the simple situation considered in this appendix, we cannot derive the linear relation between $\lambda$ and $a$ analytically.

\label{lastpage}
\end{document}